\documentstyle[aps]{revtex}
\begin{document}
\draft
\title{Femtoscopy with unlike particles}
\author{R.~Lednick\'{y}}
\maketitle
\maketitle
\begin{center}
{\small
{\it
Institute of Physics ASCR,
Na Slovance 2, 18221 Prague 8, Czech Republic.\\
Max-Planck-Institute f\"ur Physik,
F\"ohringer Ring 6, D-80805 Munich, Germany,\\
SUBATECH, 
4, rue Alfred Kastler, F-44070 Nantes Cedex 03, France.
}}
\end{center}

\begin{abstract}
The possibilities of unlike particle correlations for a study
of space--time characteristics of particle production are
demonstrated. 
The correlation data from heavy ion collisions is discussed,
particularly - in terms 
of the transport RQMD and hydrodynamic models.
An attention is paid to the relative space-time asymmetries 
in the production of different particle species,
measured as a ratio of the correlation functions corresponding to
different directional selections of the relative momentum vector.
Being sensitive to the relative time delays and
to the intensity of collective flows,
these asymmetries can become a promising tool
in a study of the deconfinement transition and 
the formation of quark-gluon plasma.
\end{abstract}

\section{Introduction}
                                                                     
The momentum correlations of particles at small relative velocities
are widely used to study space-time characteristics of the
production processes, so serving as a correlation femtoscope.
Particularly, for non-interacting identical particles, like photons, 
this technique is called intensity or particle interferometry.
In this case the correlations appear solely due to the effect of 
quantum statistics (QS) \cite{GGLP60,KP72}.
This effect has a deep analogy in astronomy \cite{KP75}, where it 
leads to the two--photon space--time correlations and allows one
to measure the angular radii of stars by studying the dependence 
of the two-photon coincidence rate on the distance between the detectors
(HBT effect \cite{hbt}).
In particle physics, the QS interference was first
observed as an enhanced production of the
pairs of identical pions with small opening angles 
(GGLP effect \cite{GGLP60}). Later on, similar to astronomy,
Kopylov and Podgoretsky \cite{KP72}
suggested to study the interference effect in terms of the correlation
function.\footnote
{Note that though both the KP and HBT methods are based on 
the QS interference, they represent just orthogonal measurements \cite{KP75}.
The former, being the momentum-energy measurement, yields the
space-time picture of the source, while the latter does the opposite.
In particular, the HBT method provides the information on the 
characteristic relative three-momenta of the emitted photons and so,
when divided by the mean detected momentum, on the angular
size of a star but, of course, - no information on the star radius or its
lifetime. 
}  
In a series of papers, they settled the basics
of correlation femtometry. Particularly, they
clarified the role of the space--time parameters studying 
their effect on the directional and velocity dependence
of the correlation function in various physical situations. 

The effect of QS is usually considered in the limit of a low
phase-space density such that the possible multi-particle effects can be
neglected.
This approximation seems to be justified by present experimental
data which does not point to any spectacular multi-boson effects neither
in single-boson spectra nor in two-boson correlations.
These effects can however clearly manifest themselves in some rare events
({e.g.}, those with large pion multiplicities) or in
the eventually overpopulated regions of momentum
space; see, {e.g.}, \cite{pra93,led00} and 
references therein.

The momentum correlations of particles emitted at nuclear distances
are also influenced by the effect of particle interaction 
in the final state (FSI) \cite{koo,ll1}.\footnote
{This effect has no analogy in the space--time correlations in
astronomy, where the particles are emitted and detected at
macroscopic distances and so, their mutual interaction is absent
in principle.
}
It should be emphasised that, depending on the characteristic
space--time separation of the particle emitters, both the
Coulomb and strong FSI can significantly influence the shape
of the correlation function. 
Thus the effect of the Coulomb interaction dominates the correlations
of charged particles at very small relative momenta
(of the order of the inverse Bohr radius of the two-particle system), 
respectively suppressing or 
enhancing the production of particles with like or unlike charges.
As a result, the correlation function of two charged particles
emitted at large relative distances in their c.m.s. is mainly
determined by the Coulomb interaction; it is increasingly sensitive
to these distances with the increasing particle masses and charges,
i.e. with the decreasing Bohr radius $|a|$ of the pair.
Regarding the effect of the strong FSI, it is 
quite small for two pions, while for two nucleons it is often a dominant one
due to the very large magnitude of the s-wave singlet scattering length
of about 20 fm.
Though the FSI effect complicates the correlation analysis,
it is an important source of information allowing to
\begin{itemize}
\item{perform the correlation femtometry with unlike particles
\cite{ll1,BS86};
}
\item{get information on the strong scattering amplitudes 
that are hardly accessible by other means;}
\item{access the relative space--time asymmetries in particle
production (e.g., relative time delays)
studying directional asymmetries of the unlike
particle correlations \cite{LLEN95}.
}
\end{itemize}

\section{Formalism}
Following Kopylov and Podgoretsky, we define the two-particle
correlation function ${\cal R}(p_{1},p_{2})$ as a
ratio of the differential two-particle production cross section to the
reference one which would be observed in the absence of the
effects of QS and FSI. 
In heavy ion collisions, one 
can neglect kinematic constraints
and most of the dynamical correlations and construct the
reference distribution by mixing the particles from different
events, normalising the correlation function to unity at sufficiently
large relative velocities. 
As usual, we assume
that the correlation of two particles
emitted with a small relative velocity 
is influenced by the effects of
their mutual QS and FSI only\footnote
{Besides the events with a large phase-space density fluctuations,
this assumption may be not justified also in low energy 
heavy ion reactions
when the particles are produced in a strong Coulomb field of residual
nuclei. To deal with this field a quantum adiabatic (factorisation)
approach can be used \cite{3body}.
} 
and that the momentum dependence of the one-particle emission probabilities
is inessential when varying the particle four-momenta 
$p_{1}$ and $p_{2}$ by the
amount characteristic for the correlation due to QS and FSI
({\it smoothness assumption}). Clearly, the latter assumption, requiring
the components of the mean space-time distance
between particle emitters 
much larger than those of the space-time extent
of the emitters, is well justified for heavy ion collisions.

The correlation function is then given by a square of the properly
symmetrized Bethe-Salpeter
amplitude in the continuous spectrum of the two-particle states,
averaged over the four-coordinates $x_{i}=\{t_{i},{\bf r}_{i}\}$
of the emitters and over the total spin $S$ of the two--particle
system \cite{ll1}.
After the separation of the unimportant phase factor 
due to the c.m.s. motion, 
this amplitude depends only on the relative four-coordinate 
$x\equiv \{t,{\bf r}\}=x_1-x_2$ and the generalised relative momentum 
$\widetilde{q}=q-P(qP)/P^2$, where $q = p_1-p_2$, 
$P=p_1+p_2$ and $qP = m_1{}^2-m_2{}^2$;
in the two-particle c.m.s., ${\bf P} = 0$, 
$\tilde q = \{0,2{\bf k}^*\}$ and $x = \{t^*,{\bf r}^*\}$.
At equal emission times of the two particles in their c.m.s.
($t^* \equiv t_1^*-t_2^* = 0$),
the latter amplitude coincides with 
a stationary solution $\psi ^{S}_{-{\bf k}^*}({\bf r}^*)$ of the 
scattering problem having at large distances
${r}^*$ the asymptotic form of a 
superposition of the plane and outgoing spherical waves 
(the minus sign of the vector ${\bf k}^{*}$ corresponds to the reverse
in time direction of the emission process). 
The Bethe-Salpeter amplitude can be usually substituted  by
this solution ({\it equal time} approximation),\footnote
{For non--interacting particles, the non--symmetrized Bethe-Salpeter 
amplitude reduces to the plane wave 
${\rm e}^{i\tilde qx/2}\equiv {\rm e}^{-i{\bf k}^*{\bf r}^*}$
which is independent of the relative time in the two--particle c.m.s. 
and so, coincides with the corresponding equal--time amplitude.
For interacting particles, the {\it equal time} approximation is valid
on condition \cite{ll1}
$ |t^*|\ll m_{2,1}r^{*2}$ for 
${\rm sign}(t^*)=\pm 1$ respectively.
This condition is usually satisfied
for heavy particles like kaons or
nucleons. But even for pions, the $t^{*}=0$ approximation
merely leads to a slight overestimation (typically $<5\%$) of the strong
FSI effect and, 
it doesn't influence the leading zero--distance 
($r^{*}\ll |a|$) effect of the Coulomb FSI.}
so that
$
{\cal R}(p_{1},p_{2})\doteq
\sum_{S}\rho_{S}
\langle |\psi_{-{\bf k}^{*}}^{S(+)}({\bf r}^{*})|^{2}
\rangle _{S}.
$
Here the averaging is done over the four--coordinates of the emitters
at a given total spin $S$ of the two--particles,
$\rho_{S}$ is the corresponding population probability,
$\sum_{S}\rho_{S} = 1$.
{
For unpolarised particles
with spins $s_{1}$ and $s_{2}$ the probability 
$\rho_{S}=(2S+1)/[(2s_{1}+1)(2s_{2}+1)]$.
Generally, the correlation function is sensitive to particle
polarisation. For example, if two spin-1/2 particles are emitted with 
polarisations $\mbox{\boldmath${\cal P}$}_1$ and 
$\mbox{\boldmath${\cal P}$}_2$ then 
\cite{ll1} $\rho_0=(1-\mbox{\boldmath${\cal P}$}_1\cdot
\mbox{\boldmath${\cal P}$}_2)/4$ and $\rho_1=(3+
\mbox{\boldmath${\cal P}$}_1\cdot\mbox{\boldmath${\cal P}$}_2)/4$. }

For non-interacting identical particles, the QS 
symmetrization:
$
\psi_{-{\bf k}^{*}}^{S(+)}({\bf r}^{*}) \rightarrow
[\psi_{-{\bf k}^{*}}^{S(+)}({\bf r}^{*})+(-1)^{S}
\psi_{{\bf k}^{*}}^{S(+)}({\bf r}^{*})]/\sqrt{2}
$ 
leads to
the characteristic feature of the correlation function - the
presence of the interference maximum or minimum at small
relative momenta
$|{\bf q}|=|{\bf p}_1-{\bf p}_2|$ 
with the width reflecting the inverse space-time extent 
of the production region.
For example, assuming that for a  
fraction $\lambda$ of the pairs, 
the particles are emitted independently according to 
one--particle amplitudes of a Gaussian form characterised by the
space--time dispersions $r_0^2$ and $\tau_0^2$ 
while, for the remaining fraction $(1-\lambda)$ 
related to very long--lived sources ($\eta$, $\eta'$, $K^0_s$, 
$\Lambda$, \dots), the relative distances $r^*$ between the emitters
in the pair c.m.s. are extremely large, 
one has
$
{\cal R}(p_{1},p_{2})= 1+\lambda\sum_S{\rho}_S(-1)^S\exp(-r_0{}^2{\bf
q}^2-\tau_0{}^2q_0^2),
$
where 
$\sum_S {\rho}_S(-1)^S=(-1)^{2s}/(2s+1)$ for
initially unpolarised spin--$s$ particles.

One may see that, due to the relation $q_0 = {\bf v}{\bf q}$
(following from the equality $qP=0$), strongly correlating
the energy difference $q_0$ with the projection of the 
three--momentum difference ${\bf q}$ on the direction of the pair
velocity ${\bf v}={\bf P}/P_0$,
the correlation function at $v\tau_{0} > r_{0}$ substantially depends
on the direction of the vector  ${\bf q}$
even in the case of a
spherically symmetric spatial form of the production region.
Generally, the directional and velocity dependence of the correlation
function can be used to determine both
the duration of the emission process and the form
of the emission region \cite{KP72}, as well as - to reveal 
the details of the
production dynamics (such as collective flows; 
see, {e.g.} \cite{PRA84}).
For this, the correlation functions are often analysed
in terms of the {out} (x), {side} (y) and {longitudinal} (z)
components of the relative momentum vector ${\bf q}=\{q_x,q_y,q_z\}$
\cite{pod83,ber94};
the {out} and {side} denote the transverse,
with respect to the reaction axis, components
of the vector ${\bf q}$, the {out} direction is
parallel to the transverse component of the pair three--momentum.
Note that for unlike particles, the relative momentum $q$ has to be
substituted by the generalised one, $\widetilde{q}$, vanishing
at equal particle velocities. 

It is well known that particle correlations at high energies
usually measure only a small part of the space-time emission volume,
being only slightly sensitive to its increase related to the fast 
longitudinal motion of the particle sources. In fact, 
due to limited source decay momenta 
${\rm p}^{(s)}$ 
of few hundred MeV/c, the correlated particles with nearby velocities
are emitted by almost comoving sources and so - at nearby space--time
points. 
In other words, the maximal contribution of the relative motion
to the correlation radii in the two--particle c.m.s. is limited
by the moderate source decay length $\tau {\rm p}^{(s)}/m$.
The dynamical examples are sources-resonances,
colour strings or hydrodynamic expansion.
To substantially eliminate the effect of the longitudinal motion, 
the correlations can be
analysed in terms of the invariant variable
$q_{inv}\equiv Q = (-\widetilde{{\bf q}}^2)^{1/2} = 2k^*$ and
the components of the  momentum difference in pair c.m.s.
(${\bf q}^*\equiv {\bf Q}= 2{\bf k}^*$) or in
the longitudinally comoving system (LCMS) \cite{cso91}.
In LCMS each pair is emitted transverse to the reaction axis
so that the generalised relative momentum
$\widetilde{{\bf q}}$ coincides with ${\bf q}^*$ 
except for the component
$\widetilde{q}_x=\gamma_{t}q_x^*$,
where $\gamma_{t}$ is the LCMS Lorentz factor of the pair.

\section{Femtometry with unlike particles}
The complicated dynamics of particle production,
including resonance decays and particle rescatterings,
leads to essentially non--Gaussian tail of the
$r^*$--distribution. 
Therefore, due to different $r^*$--sensitivity
of the QS, strong and Coulomb FSI effects,
one has to be careful when analysing the correlation functions
in terms of simple models. 
Thus, the QS and strong FSI effects are influenced by the $r^*$--tail
mainly through the suppression parameter $\lambda$ while,
the Coulomb FSI is sensitive to the distances as large as
the pair Bohr radius $|a|$.
For $\pi\pi$, $\pi K$, $\pi p$, $KK$, $K p$ and $p p$ pairs, 
$|a|=$ 387.5, 248.6, 222.5, 109.6, 83.6 and 57.6 fm, 
respectively.
Clearly, the usual Gaussian parametrisations of the components
of the vector ${\bf r}^*$ may happen to be insufficient in the
case of charged particle correlations, leading to 
inconsistencies in the treatment of QS and FSI effects
(the Coulomb FSI contribution requiring larger effective radii).
These problems can be at least partially overcome with the help of 
transport code simulations accounting for the dynamical evolution
of the emission process and providing the phase space information
required to calculate the QS and FSI effects on the correlation function.

Thus, in a preliminary analysis of the NA49 correlation data from 
central $Pb+Pb$ 158 AGeV collisions \cite{lna49}, 
we have used the events simulated
with the RQMD v.2.3 code \cite{rqmd}. The correlation functions 
have been calculated using the code of Ref. \cite{ll1},
weighting the simulated pairs by squares of the
corresponding wave functions. 
To account for a possible
mismatch in $\langle r^*\rangle$, the RQMD simulations
have been done with the space--time coordinates of the emission points
scaled by a factor of 0.7, 0.8 and 1. The $Q$--dependence of the
correlation function was than fitted according to the formula \cite{lna49}
$
{\cal R}(Q)= {\rm norm}\cdot [{\rm purity}\cdot 
{\rm RQMD}(r^*\rightarrow {\rm scale}\cdot r^*)+ (1-{\rm purity})],
$
where the dependence on the scale parameter has been introduced
using the quadratic interpolation of the points simulated at the three
different scales. Based on a sample of 900k events,
representing a quarter of the available statistics,
the parameters fitted for the $\pi^+\pi^-$, $\pi^+ p$ and $\pi^- p$ 
systems are respectively: 
purity $= 0.77\pm 0.01$, $0.76\pm 0.03$ and $0.74\pm 0.04$,
scale $= 0.78\pm 0.01$, $0.87\pm 0.03$ and $0.88\pm 0.04$.
The values of the purity parameter are in reasonable agreement 
with the expected contamination
from strange particle decays and particle misidentification.
The fitted values of the scale parameter indicate that RQMD
overestimates the distances $r^*$ by 10-20$\%$. Similar overestimation
has been also observed when comparing RQMD predictions with the 
NA49 data on $pp$ and $\pi^\pm\pi^\pm$ correlations
\cite{ppna49,gan99,pipina49}.
Regarding the absolute estimates of $\langle r^*\rangle$,
they depend on the truncation of large distances. 
For example, for $\pi^+\pi^-$ system, 
RQMD model yields 21 and 29 fm for the means
truncated at 50 and 500 fm respectively. 
Scaling this numbers down by a factor of 0.78,
one gets for the corresponding experimental truncated means
16.4 and 22.6 fm for $r^* <$ 39 and 390 fm respectively.   
 
Recently, there appeared data on $p\Lambda$ correlation functions
from $Au+Au$ experiment E985 at AGS \cite{lis01}. 
As the Coulomb FSI is absent
in this system, one avoids here the problem of its sensitivity
to the $r^*$--tail. Also, the absence of the Coulomb suppression
of small relative momenta makes this system more sensitive to the
radius parameters as compared with $pp$ correlations \cite{wan99}.
Despite of rather large statistical errors,
a significant enhancement is seen at low relative momentum,
consistent with the known positive singlet and triplet 
$p\Lambda$ s--wave scattering lengths (defined here as the 
corresponding amplitudes at $k^*=0$) of 2.3 and 1.8 fm respectively;
the corresponding effective radii are 3.0 and 3.2 fm.
In fact, using the analytical expression for the correlation
function from Ref. \cite{ll1} (originally derived for $pn$ system),
one gets a good fit of
the combined (4, 6 and 8 AGeV) correlation function
with the suppression parameter $\lambda= 0.5\pm 0.2$ and
the Gaussian radius $r_0= 4.5 \pm 0.7$ fm.
The $\lambda$--value is in agreement with the product of the
reconstruction $\Lambda$ purity ($\sim 80\%$) and the non--feed--down
purity ($70-75\%$) (the feed-down from $\Sigma^0$s is estimated
at $25-30\%$). As for the fitted radius, it agrees with the radii
of 3-4 fm obtained from $pp$ correlations in 
heavy ion collisions at GSI, AGS and SPS energies.

There is now coming a high quality data on particle correlations
from experiment STAR at RHIC \cite{ret01}. Particularly, the
$\pi K$ correlations in all possible charge states are available
from $Au+Au$ collisions at 130 AGeV c.m.s. energy.
They practically coincide for the same and opposite charges
and show the mirror symmetry. This points to the
same production mechanism of positive and negative pions (kaons).
A preliminary analysis of this data within a static sphere model
yields the radius of about 7 fm \cite{ret01}.

\section{Accessing scattering amplitudes}
In case of a poor knowledge of the two--particle strong interaction,
which is the case for {\it exotic} systems like ($M=$ meson)
$MM$, $M \Lambda$ or $\Lambda \Lambda$, 
the correlation measurements can be also used to study the latter. 
Particularly important is the experimental study of $\Lambda\Lambda$ 
interaction, especially in view of a recent experimental 
indication on the enhanced $\Lambda\Lambda$ production near 
threshold \cite{ahn98} 
and its possible connection with  the 6-quark H dibaryon problem. 

On the absence of a noticeable FSI, the behaviour of the 
correlation function ${\cal R}$ of two identical particles 
at small relative momenta $Q=2k^*$ in pair rest frame is determined by 
the symmetry requirement of QS: at $Q\rightarrow 0$ 
the contribution of the even total spin $S$ ({e.g.}, 
singlet part ${\cal R}_s$) is enhanced by a factor of 2 
while that of the odd $S$ ({e.g.}, the triplet part 
${\cal R}_t$) vanishes; 
the enhancement or dip width is inversely 
related to the effective radius $r_0$ of the emission region. 
In heavy ion collisions, $r_0$ can be 
considered much larger than the range $d$ of the 
strong interaction potential. 
The FSI contribution is then independent of the actual 
potential form \cite{gkll86}. At small $Q$, it is determined by the s-wave 
scattering amplitudes $f^S(k^*)$ 
\cite{ll1}. 
In case of $|f^S|>r_0$, this contribution is of the order of 
$|f^S/r_0|^2$ and dominates over the effect of QS. In the opposite case,
the sensitivity of the correlation function to the scattering
amplitude is determined by the linear term $f^S/r_0$.

To demonstrate the possibilities of the correlation measurements
of the scattering amplitudes, we have repeated the analysis of the
NA49 $\pi^+\pi^-$ correlation function within the RQMD model,
introducing the additional parameter sisca (strong interaction scale),
redefining the original scattering length $f_0=$ 0.232 fm:
$f_0\rightarrow {\rm sisca}\cdot f_0$.
The introduction of this new scale lead to a substantial improvement
of the fit quality and to a noticeable increase and decrease of the 
purity and $r^*$--scale parameters respectively: 
purity $= 0.81\pm 0.01$, scale $=0.72\pm 0.02$.
The parameter sisca $=0.63\pm 0.08$ appears to be significantly
lower than unity, showing that the correlation data prefer the
value of the s-wave $\pi^+\pi^-$ scattering length 
$f_0=2a_0^0+a_0^2$ ($a_0^{0,2}$ are the two--pion isosinglet and isotensor
s-wave scattering lengths) by $\sim 30\%$ lower
than than the present table value. To a similar shift ($\sim 20\%$) 
point also the recent BNL data on $K_{l4}$ decays \cite{pis01}.
These results are in agreement with the two--loop calculation 
in the chiral perturbation theory with a standard value of the
quark condensate \cite{col00}.
Comparing with the theoretical predictions, one should have in mind
that they are subject to the electro-magnetic corrections on the level
of several percent and that the correlation measurement underestimates
$f_0$ by a few percent due to the use of the equal--time
approximation.

As for the $\Lambda\Lambda$ system, one can try
to estimate the singlet 
$\Lambda\Lambda$ s--wave scattering length $f_0$, fitting
the recent NA49 data on
$\Lambda\Lambda$ correlations in $Pb+Pb$ collisions at
158 AGeV \cite{blu01}.
Using the analytical expression for the correlation
function from Ref. \cite{led99} 
(originally derived for $nn$ system \cite{ll1}),
one gets $\lambda= 0.9\pm 0.6$, $r_0= 1.5\pm 0.3$ fm and
$f_0= 0.1\pm 0.5$ fm. Fixing the suppression parameter $\lambda$
at a reasonable value of 0.5, one gets $r_0= 1.8\pm 0.3$ fm and
$f_0= -2.6\pm 2.6$ fm.
Though the fitting results are not very restrictive, they 
certainly exclude
the possibility of a large positive singlet scattering length 
comparable to that for the two--nucleon system.

\section{Accessing relative space-time asymmetries}

The correlation function of two non--identical particles, 
compared with the identical ones,
contains a principally new piece of information on the relative
space-time asymmetries in particle emission \cite{LLEN95}.
Since this information enters in the two-particle amplitude
$\psi_{-{\bf k}^{*}}^{S(+)}({\bf r}^{*})$
through the terms odd in ${\bf k}^{*}{\bf r}^{*}$,
it can be accessed studying the correlation functions
${\cal R}_{+i}$ and ${\cal R}_{-i}$ 
with positive and negative projection $k^*_i$ on
a given direction ${\bf i}$ or, - the
ratio ${\cal R}_{+i}/{\cal R}_{-i}$.
For example, ${\bf i}$ can be the direction of the pair
velocity or, any of the out (x), side (y), longitudinal (z)
directions. Note that in the LCMS system, one has 
$r^*_i=r_i$ except for 
$r_x^*\equiv\Delta x^*=\gamma_{t}(\Delta x-v_{t}\Delta t)$,
where $\gamma_{t}=(1-v_t{}^2)^{1/2}$ and $v_t=P_t/P_0$
are the pair LCMS Lorentz factor and velocity.
One may see that the asymmetry in the out (x) direction
depends on both space and time asymmetries 
$\langle\Delta x\rangle$ and $\langle\Delta t\rangle$.
In case of a dominant Coulomb FSI, the intercept of the correlation
function ratio is directly related with the asymmetry 
$\langle r^*_i\rangle$:
$
{\cal R}_{+i}/{\cal R}_{-i}\approx 1+
2\langle r_i^*\rangle /a, 
$ 
where $a=(\mu z_{1}z_{2}e^{2})^{-1}$ is the Bohr radius of the
two-particle system taking into account the sign of the interaction
($z_ie$ are the particle electric charges,
$\mu$ is their reduced mass).

At low energies, the particles in heavy ion collisions are emitted
with the characteristic emission times of tens to hundreds fm/c so that
the observable time shifts should be of the same order \cite{LLEN95}.
Such shifts have been indeed observed with the help of the 
${\cal R}_{+}/{\cal R}_{-}$ correlation ratios for 
proton-deuteron systems in several heavy ion
experiments at GANIL \cite{ghi95} indicating,
in agreement with the coalescence model, that deuterons are on average 
emitted earlier than protons.

For ultra-relativistic heavy ion collisions,
the sensitivity of the ${\cal R}_+/{\cal R}_-$ correlation ratio to the 
relative time shift 
$\langle\Delta t\rangle$ (introduced {\it ad hoc}) 
was studied for various two-particle systems simulated using the
transport codes \cite{ALICE}.
The scaling of the effect with the space-time asymmetry and 
with the inverse Bohr radius $a$ was clearly illustrated.            
It was concluded that 
the ${\cal R}_{+}/{\cal R}_{-}$ ratio can be sensitive to the shifts 
in the particle emission times of the order of a few fm/c. 
Motivated by this result, the correlation asymmetry 
for the $K^+K^-$ system has been studied
in a two-phase thermodynamic evolution model and
the sensitivity has been demonstrated 
to the production of the transient strange quark matter state
even if it decays on strong interaction time scales \cite{sof97}.
The method sensitivity to the space-time asymmetries arising also in the
usual multi-particle production scenarios was demonstrated
for AGS and SPS energies using the transport code RQMD 
\cite{lna49,lpx,vol97}. 
At AGS energy, the $Au+Au$ collisions have been simulated and 
the $\pi p$ correlations have been studied
in the projectile fragmentation region where proton directed
flow is most pronounced and where the proton and pion sources
are expected to be shifted 
relative to each other both in the longitudinal 
and in the transverse directions
in the reaction plane.
It was demonstrated \cite{vol97} that 
the corresponding ${\cal R}_{+}/{\cal R}_{-}$ ratios are
sufficiently sensitive to reveal the
shifts; they were confirmed in the directional analysis of
the experimental AGS correlation data \cite{mis98}.

At SPS energy, the simulated central $Pb+Pb$ collisions 
yield practically zero asymmetries for $\pi^+\pi^-$ system
while, for $\pi^\pm p$ systems, the LCMS $x$- and $t$-asymmetries 
are $\langle\Delta x\rangle = -6.2$ fm,
$\langle\Delta t\rangle = -0.5$ fm/c, 
$\langle\Delta x^*\rangle = -7.9$ fm in the central rapidity
window \cite{lpx} and,
$\langle\Delta x\rangle = -5.2$ fm,
$\langle\Delta t\rangle = 2.9$ fm/c, 
$\langle\Delta x^*\rangle = -8.5$,
for the NA49 acceptance (shifting the rapidities into the
forward hemisphere).
Besides, $\langle x\rangle$ increases with particle $p_t$ or
$u_t=p_t/m$, starting from zero due to kinematic reasons.
The asymmetry arises because of a faster increase with $u_t$
for heavier particle. 
The non--zero positive value of 
$\langle x\rangle=\langle {\bf r}_t\hat{\bf x}\rangle$
($\hat{\bf x}={\bf p}_t/p_t$ and ${\bf r}_t$ is the transverse
radius vector of the emitter) and the hierarchy
$\langle x_\pi\rangle<\langle x_K\rangle<\langle x_p\rangle$
is a signal of a universal transversal collective flow.
To see this, one should take into account that the mean thermal
velocity $\langle\beta^T\rangle$ as well as the mean particle transverse
velocity $\langle\beta_t\rangle$ are smaller for heavier particles.
Thus, in the non--relativistic approximation,
one has $\mbox{\boldmath$\beta$}_t\doteq\mbox{\boldmath$\beta$}_0+
\mbox{\boldmath$\beta$}_t^T$ and 
$\langle x\rangle=
\langle r_t(\beta_0+\beta_t^T\cos\phi)/\beta_t\rangle
\approx \langle r_t\beta_0/\beta_t\rangle$,
where $\mbox{\boldmath$\beta$}_0\parallel {\bf r}_t$ is the
transversal flow velocity, $\phi$ is the angle between the
vectors $\mbox{\boldmath$\beta$}_t^T$ and $\mbox{\boldmath$\beta$}_0$. 
As a result, in case of a locally equilibrated expansion process,
one expects a negative asymmetry $\langle x\rangle\equiv
\langle x_1-x_2\rangle$ provided $m_1<m_2$. Moreover, this asymmetry
vanishes in both limiting cases: $\beta_0\ll \beta^T$ and
$\beta_0\gg \beta^T$. 
These conclusions agree with the calculations
in the longitudinal-boost invariant 
hydrodynamic model. Thus, assuming the linear transversal
rapidity profile with $y_t=0.4$ at the characteristic Gaussian 
transverse radius of 6 fm and the temperature of 140 MeV
(corresponding to central $Pb+Pb$ collisions at SPS energy) and,
using Eq.~(30) of Ref.~\cite{akk96} to calculate $\langle x\rangle$
as a function of $u_T$,
one confirms a faster rise of $\langle x\rangle$ for heavier particles
and gets $\langle \Delta x\rangle= -3$ fm and 
$\langle \Delta x^*\rangle= -4$ fm for $\pi^\pm p$ systems.
In fact, the NA49 data on ${\cal R}_{+x}/{\cal R}_{-x}$ ratio
for $\pi^+p$ and $\pi^-p$ systems show consistent mirror symmetric
deviations from unity, their size of several percent and the
$Q$--dependence being in agreement with RQMD calculations corrected
for the resolution and purity. The predictions of the simple
hydrodynamic model (assuming the universal freeze-out time)
appear to be $\sim 50\%$ too low.

Similar pattern of the correlation asymmetries has been reported
also for $\pi^\pm K^\pm$ and $\pi^\pm K^\mp$ correlation function
ratios in the out (x) direction from experiment STAR at RHIC \cite{ret01}. 
They seem to be in agreement with the hydrodynamic type calculations
without any corrections, thus leaving a room for the additional
space--time asymmetries (e.g., due to the time shifts).

\section{Conclusions}
Analysing the experimental and simulated correlation data from
heavy ion collisions,
we have shown that unlike particle correlations can serve
as an important femtometry tool complementary to the usual
interferometry with identical particles.
We have demonstrated that two--particle correlations provide
a useful information on the strong scattering amplitudes  
that are hardly accessible by other means.
We have shown that directional asymmetries of unlike particle
correlations contain a principally new piece of information on the relative
space-time asymmetries in particle emission.
The data on correlation asymmetries in heavy ion collisions at
AGS, SPS and RHIC appear to be in quantitative or qualitative
agreement with transport or hydrodynamic calculations.
Being sensitive to relative time delays and collective flows,
the correlation asymmetries can be useful to study 
the effects of the quark--gluon
plasma phase transition.
As for the detection of the unlike particles with close velocities, 
there is practically no problem with the two-track resolution since these
particles,
having either different momenta or different charge-to mass ratios,
have well separated trajectories in the detector magnetic field.
For the same reason, however, a large momentum acceptance of the
detector is required.

This work was supported by 
GA Czech Republic, Grant No. 202/01/0779.

\end{document}